\documentclass[12pt]{article}

\usepackage{amssymb}
\usepackage{amsmath}
\usepackage{latexsym}
\usepackage{yfonts}

\catcode`@=11
\renewcommand{\section}{\@startsection{section}{1}{0pt}{\medskipamount}
{\medskipamount}{\large\bf}} \numberwithin{equation}{section}
\catcode`@=12

\def\beq{\begin{eqnarray}}    
\def\eeq{\end{eqnarray}}      

\def\ln{\,\mbox{ln}\,}                  
\def\Box{\square}                       

\def\pa{\partial}                       

\def\={\ =\ }

\begin{document}

\begin{center}

{\Large\bf Gauge dependence of alternative flow equation for the functional
renormalization group}

\vspace{18mm}

{\large
Peter M. Lavrov$^{(a, b)}\footnote{E-mail:
lavrov@tspu.edu.ru}$\; }

\vspace{8mm}

\noindent  ${{}^{(a)}} ${\em
Tomsk State Pedagogical University,\\
Kievskaya St.\ 60, 634061 Tomsk, Russia}

\noindent  ${{}^{(b)}} ${\em
National Research Tomsk State  University,\\
Lenin Av.\ 36, 634050 Tomsk, Russia}

\vspace{20mm}

\begin{abstract}
\noindent
The gauge dependence problem of alternative flow equation for the functional
renormalization group is studied. It is shown that the effective two-particle irreducible
effective action depends on gauges at any value of IR parameter $k$. The situation
with gauge dependence is similar to the standard formulation based on the
effective one-particle irreducible average action.

\end{abstract}

\end{center}

\vfill

\\

\newpage

\section{Introduction}
\noindent
Recently \cite{Lav1} it has been proved that in the case of gauge theories
the effective average action of the functional renormalization group (FRG)
\cite{Wet1,Wet2} found as a solution to the flow equation depends on
gauges at any
value of IR cutoff $k$ that makes impossible a physical interpretation
of results obtained. The main reason of this situation is related
to the fact that
standard formulation of the FRG violates the gauge invariance of an initial
classical action and  the BRST symmetry \cite{brs1,t} of quantum action.
In turn within standard perturbation approach to Quantum Field Theory
it leads to the gauge dependence of effective average action even on-shell
in contrast with corresponding property of effective action constructed according to  the
Faddeev-Popov rules \cite{FP}.
In this connection it is necessary to mention a redefinition
of standard FRG approach in the form respecting the BRST symmetry
\cite{Morris1,IIS,Morris2}. It was proposed to arrive at this nice feature
in contrast to the standard FRG approach with
the help of regularization of an initial classical action which remains
gauge invariant. Realization of this program in practice meets with serious
difficulties \cite{Lav2}. Strongly speaking up to now a consistent
and constructive procedure to realize explicitly the program
does not exist because in \cite{Morris1} it was
claimed a possibility to construct a regularized gauge
invariant classical action perturbatively and in \cite{IIS}
it was just assumed existence of regularized interaction action.

Some years ago it was proposed \cite{LSh} to realize main ideas of
the FRG approach using concept of effective action with composite operators
or, in another words, effective two-particle irreducible (2PI)  action introduced
in paper \cite{CJT} and then studied for gauge theories from point of view the
gauge dependence in \cite{Lav-tmph}. Quite recently
the alternative flow equation using slightly different way in comparison with
\cite{LSh} and \cite{CJT} has been introduced \cite{AMNS-1}.
Main difference is related
with introduction of external sources when they are considered as independent
variables \cite{CJT} or they depend on the IR parameter $k$ of FRG \cite{AMNS-1}.

In standard perturbation
approach to quantum theory of gauge fields
the effective actions with composite operators used in
\cite{LSh,AMNS-1} are gauge independent on-shell. But by itself
this property does not guarantee the gauge independence of
the effective action with composite operators found
as non-perturbative solution to the flow equation as  it was shown \cite{Lav1} in the case
of standard FRG approach. In the present paper we are going
to analyze the derivation and the gauge dependence of alternative flow equations
for the FRG appearing in two ways of introduction of external sources.

The paper is organized as follows. In Sec. 2 basic properties and gauge dependence
of effective action with composite operators in the framework
of standard perturbation approach to gauge theories are considered.
In Sec. 3  the derivation of alternative flow equations for the
effective action with composite operators is given.
In Sec. 4 the gauge dependence of the alternative flow equations
is investigated.
Finally, in Sec. 5 the results obtained in the paper are discussed.

We use the DeWitt's condensed notations \cite{DeWitt}.
We employ the notation $\varepsilon(A)$ for the Grassmann parity of
any quantity $A$.  The functional derivatives with respect to fields and sources
are considered as right and left correspondingly.   The left
functional derivatives with respect to
fields  are marked by special symbol $"\rightarrow"$.
Arguments of any functional are enclosed in square brackets
$[\;]$,
and arguments of any function are enclosed in parentheses, $(\;)$.
The symbol $F_{,A}[\phi,...]$ means the
right derivative of $F[\phi,...]$ with respect to field $\phi^A$.

\section{Ward identity and gauge dependence of 2PI effective  action}
\noindent
Let us start from some initial classical action $S_0[A]$
of the fields $A^i$, with Grassmann parities
$\varepsilon(A^i)\equiv\varepsilon_i$, being invariant under the
gauge transformations ($X_{,i}\equiv\delta X/\delta A^i$)
\begin{eqnarray}
\label{v1}
\delta A^i=R^i_{\alpha}(A)\xi^{\alpha},\quad
S_{0,i}[A]R^i_{\alpha}(A) =0,
\end{eqnarray}
where $\xi^{\alpha}$ are arbitrary functions with Grassmann parities
$\varepsilon(\xi^{\alpha})\equiv\varepsilon_{\alpha}$,
$\alpha=1,2,...,m$, and $R^i_{\alpha}(A)$,
$\varepsilon(R^i_{\alpha}(A))=\varepsilon_i + \varepsilon_{\alpha}$
are generators of gauge transformations.
We restrict yourself by  the case of Yang-Mills type of gauge theories when the
 algebra
of generators $R^i_{\alpha}(A)$ has the form:
\begin{eqnarray}
\label{v2}
R^i_{\alpha ,
j}(A)R^j_{\beta}(A)-(-1)^{\varepsilon_{\alpha}\varepsilon_
{\beta}}R^i_{\beta ,j}(A)R^j_{\alpha}(A)
=-R^i_{\gamma}(A)F^{\gamma}_{\alpha\beta},
\end{eqnarray}
where $F^{\gamma}_{\alpha\beta}
=-(-1)^{\varepsilon_{\alpha}\varepsilon_{\beta}}
F^{\gamma}_{\beta\alpha}$ are  structure functions not depending on the fields $A^i$ and
the generators $R^i_{\alpha}(A)$
form a set of linear independent operators with respect to the index $\alpha$.
Standard quantization of the theory under consideration in the Faddeev-Popov method
\cite{FP}
operates with the action
\beq
\label{v3}
S_{FP}[\phi]=S_0[A]+\Psi_{,A}[\phi]R^A(\phi),
\eeq
where $\Psi[\phi]$ is a gauge fixing functional and $R^A(\phi)$,
$\varepsilon(R^A(\phi))=\varepsilon_A+1$ are the
 generators of BRST transformations of fields $\phi^A$ \cite{brs1,t},
\beq
\label{v4}
R^A(\phi)=\big(R^i_{\alpha}(A)C^{\alpha},\; 0\;,
-\frac{1}{2}(-1)^{\varepsilon_{\beta}}
F^{\alpha}_{\beta\gamma}
C^{\gamma }C^{\beta}, B^{\alpha}(-1)^{\varepsilon_{\alpha}}\big).
\eeq
In (\ref{v3}) $\phi=\{\phi^A\}$,
$\phi^A=(A^i, B^{\alpha}, C^{\alpha}, {\bar C}^{\alpha})$,
$\varepsilon(\phi^A)=\varepsilon_A$ is full set of fields in
the Faddeev-Popov quantization
with the Faddeev-Popov ghost and anti-ghost fields
$ C^{\alpha}, {\bar C}^{\alpha}$ ($\varepsilon(C^{\alpha})=
\varepsilon({\bar C}^{\alpha})=1,\;
{\rm gh}(C^{\alpha})=-{\rm gh}({\bar C}^{\alpha})=1$), respectively, and
the Nakanishi-Lautrup auxiliary fields $B^{\alpha}$
($\varepsilon(B^{\alpha})=0,\; {\rm gh}(B^{\alpha})=0$).
The action
 (\ref{v3}) is invariant under global supersymmetry (BRST symmetry) \cite{brs1,t}
\beq
\label{v5}
\delta_{B}\phi^A=R^A(\phi)\mu,\qquad
 S_{FP}[\phi]_{,A}R^A(\phi)=0,
\eeq
where  $\mu$ is a constant anti-commuting parameter.

From technical point of view the standard FRG approach involves
instead of $S_{FP}[\phi]$ the action
\beq
\label{v6}
S_{Wk}[\phi]=S_{FP}[\phi]+S_k[\phi],
\eeq
where $S_k[\phi]$ is so-called regulator action being quadratic in fields $\phi$,
\beq
\label{v7}
S_k[\phi]=\int dx \frac{1}{2}R_{k|AB}(x,y)\phi^B(y)\phi^A(x)=
\frac{1}{2}R_{k|AB}\phi^B\phi^A.
\eeq
The regulators $R_{k|AB}$ depend on IR parameter $k$ and obey the properties
\beq
\label{v8}
\lim_{k\rightarrow 0}R_{k|AB}=0,\quad
R_{k|AB}=R_{k|BA}(-1)^{\varepsilon_A\varepsilon_B}, \quad
\varepsilon(R_{k|AB})=\varepsilon_A + \varepsilon_B.
\eeq
Standard choice of $R_{k|AB}(x,y)$ in the FRG reads
\beq
\label{v9}
R_{k|AB}(x,y)=z_{AB}\frac{\Box\exp\{-\Box/k^2\}}{1-\exp\{-\Box/k^2\}}
\delta(x-y),\quad
\Box=\pa_{\mu}\pa^{\mu}, \quad z_{AB}={\rm const},
\eeq
and modifies behavior of all propagators in IR region
making finite all Feynman diagrams. Nevertheless the standard FRG cannot be considered as
a qualitative quantization method for gauge theories because at any value of the
IR parameter $k$ the effective average action depends on gauges \cite{Lav1}.
This conclusion is valid within the perturbation theory
as well as  on the level of solutions to the flow equation. Improvement of the FRG in
the form of BRST exact renormalization group \cite{Morris1} remains a big question
due to the lack of an explicit procedure for constructing
a regularized gauge-invariant classical action \cite{Lav2}.

Alternative approach to the FRG proposed in \cite{LSh} is based on idea to consider
Lagrangian density ${\cal L}_k(\phi)(x,y)$ of action (\ref{v7}),
\beq
\label{v10}
{\cal L}_k(\phi)(x,y)=\frac{1}{2}R_{k|AB}(x,y)\phi^B(y)\phi^A(x),
\eeq
as composite operator in formalism \cite{CJT} with introduction
of additional external
source $\Sigma(x,y)$ so that the generating functional of Green
functions, $Z_k[J,\Sigma]$,
is defined with the help of action $S_{FP}[\phi]+J_A\phi^A+\Sigma{\cal L}_k$.
As a profit
of this reformulation the effective action with composite operators does not depend
on gauges on-shell in the perturbation theory. Motivated by recent study \cite{AMNS-1},
here we investigate  another possibility when the regulators  $R_{k|AB}$ are considered
as sources to composite fields $\phi^A\phi^B$ but external sources $J_A$ remain
independent variables.

In what follows specific forms of the initial classical action $S_0[A]$
and the
gauge fixing functional $\Psi[\phi]$ are not essential and
we consider them in general setting. Our starting point
is definition of generating functional of Green functions with composite
operators
$Z_k=Z_k[J,R_k]$, and generating functional of connected
Green function with composite
operators, $W_k=W_k[J,R_k]$, in the form
\beq
\label{v11}
Z_{k}[J,R_k]=\int
D\phi\;\exp\Big\{\frac{i}{\hbar}\Big[S_{FP}[\phi]+J\phi+
\frac{1}{2}R_k\phi\phi \Big]\Big\}=
\exp\Big\{\frac{i}{\hbar}W_{k}[J,
R_k]\Big\}.
\eeq
Making use the change of integration variables in the form of
BRST transformations (\ref{v5}), taking into account the invariance of
$S_{FP}$ under these transformations (\ref{v5}) and triviality of corresponding
Jacobian, we derive the Ward identity
\beq
\label{v12}
\big(J_A-i\hbar R_{k|AB}\pa_{\!J_B}\big)R^A(-i\hbar\pa_{\!J})Z_{k}=0.
\eeq
In terms of functional $W_k$ the identity (\ref{v12}) rewrites as
\beq
\label{v13}
\big(J_A+ R_{k|AB}(\pa_{\!J_B}W_{k}-i\hbar \pa_{\!J_B})\big)
R^A(\pa_{\!J}W_{k}-i\hbar\pa_{\!J})\cdot 1=0.
\eeq
The 2PI effective action, $\Gamma_k[\Phi,\Delta_k]$, is introduced through the
Legendre transformation of $W_k$,
\beq
\label{v14}
\Gamma_{k}[\Phi,\Delta_k]=W_{k}[J,R_k]-J_A\Phi^A-
R_{k|AB}
\Big(\frac{1}{2}\Phi^B\Phi^A+\hbar \Delta_k^{BA}\Big),
\eeq
where
\beq
\label{v15}
\frac{\delta W_{k}}{\delta J_A}=\Phi^A,\qquad
\frac{\delta W_{k}}{\delta R_{k|AB}}=
\frac{1}{2}\Phi^B\Phi^A+\hbar \Delta_k^{BA}.
\eeq
From (\ref{v14}) and (\ref{v15}) it follows
\beq
\label{v16}
\frac{\delta \Gamma_{k}}{\delta\Phi^A}=-J_A-R_{k|AB}\Phi^B,\qquad
\frac{\delta \Gamma_{k}}{\delta \Delta_k^{AB}}=-\hbar R_{k|BA}.
\eeq
Then the Ward identity in terms of $\Gamma_k$ reads
\beq
\label{v17}
\Big(\frac{\delta\Gamma_{k}}{\delta\Phi^A}+
\frac{\delta\Gamma_{k}}{\delta(\hbar\Delta_k^{BA})}({\hat\Phi}^B-\Phi^B)
 \Big)R^A({\hat \Phi})\cdot 1=0,
\eeq
where the following notations
\beq
\label{v18}
&&{\hat \Phi}^A=\Phi^A+i\hbar \big(G_k^{'' -1}\big)^{A|{\cal A}}
\pa_{{\cal F}^{\cal A}},\\
\label{v19}
&&{\cal J}_{\cal A}=(J_A,\hbar R_{k|AB}),\quad
{\cal F}^{{\cal A}}=(\Phi^A,
\hbar\Delta_k^{AB}),\\
\label{v20}
&&(G_k^{" -1})_{{\cal A}|{\cal B}}=-
\frac{\overrightarrow{\delta}{\cal J}_{{\cal B}}({\cal
F})}{\delta {\cal F}_{{\cal A}}},\quad
(G_k^{" -1})^{{\cal A}|{\cal B}}=
-\frac{\delta{\cal F}^{{\cal B}}({\cal J})}{\delta {\cal
J}_{{\cal A}}},\\
\label{v21}
&&(G_k^{" -1})^{{\cal A}|{\cal C}}(G_k^{" -1})_{{\cal C}|{\cal B}}=
\delta^{\cal A}_{\;\;\cal B},\\
\label{v22}
&&{\cal J}_{\cal A}({\cal F})=\Big(-\frac{\delta
\Gamma_k}{\delta\Phi^A}+ \frac{1}{\hbar}\frac{\delta \Gamma_k}{\delta
\Delta_k^{AB}}\Phi^B, -\frac{\delta \Gamma_k}{\delta
\Delta_k^{AB}}\Big),\\
\label{v23}
&&{\cal F}^{\cal A}({\cal J})=\Big(\frac{\delta
W_k}{\delta J_A}, \frac{\delta W_k}{\delta R_{k|AB}}-
\frac{1}{2}\frac{\delta W_k}{\delta J_A}\frac{\delta W_k}{\delta
J_B}\Big),
\eeq
are used.

Now we want to study
the gauge dependence of functionals introduced in (\ref{v11})
and (\ref{v14}). To do this we consider infinitesimal variation
of gauge fixing functional, $\Psi[\phi]\rightarrow \Psi[\phi]+
\delta\Psi[\phi]$, and corresponding generating functional of
Green functions using a temporary designation for the functional
(\ref{v11}), $Z_{k}[J,R_k]=Z_{\Psi|k}[J,R_k]$,
\beq
\label{v24}
Z_{\Psi+\delta\Psi|k}[J,R_k]=\int
D\phi\;\exp\Big\{\frac{i}{\hbar}\Big[S_{FP}[\phi]+
\delta\Psi_{,A}[\phi]R^A(\phi)+J\phi+
\frac{1}{2}R_k\phi\phi \Big]\Big\}.
\eeq
From (\ref{v24}) it follows the variation of $Z_{\Psi|k}[J,R_k]$,
\beq
\label{v25}
\delta Z_{\Psi|k}=\frac{i}{\hbar}\delta\Psi_{,A}[-i\hbar\pa_{\!J}]
R^A(-i\hbar\pa_{\!J})Z_{\Psi|k},\quad Z_{\Psi|k}=Z_{\Psi|k}[J.R_k].
\eeq
There exists an equivalent representation of gauge dependence of $Z_{\Psi|k}$
following from the BRST symmetry of functional $S_{FP}[\phi]$
(for details see \cite{Lav1}),
\beq
\label{v26}
\delta Z_{\Psi|k}=\frac{i}{\hbar}
\big(J_A-i\hbar R_{k|AB}\pa_{\!J_B}\big)R^A(-i\hbar\pa_{\!J})
\delta\Psi[-i\hbar\pa_{\!J}]Z_{\Psi|k},
\eeq
From the equation (\ref{v26}) it follows that the vacuum functional
$Z_{\Psi|k}[0]=Z_{\Psi}[J=0,R_k=0]$
does not depend on gauges
\beq
\label{v27}
\delta Z_{\Psi|k}[0]=0.
\eeq
Due to this fact  we will omit the subscript $\Psi$ in the functionals
$Z_{\Psi|k}=Z_k$, $W_{\Psi|k}=W_k$.

The gauge dependence of functional $W_k$ is described by the equation
\beq
\label{v28}
\delta W_{k}=
\big(J_A+ R_{k|AB}(\pa_{\!J_B}W_k-i\hbar \pa_{\!J_B})\big)
R^A(\pa_{\!J}W_k-i\hbar\pa_{\!J})
\delta\Psi[\pa_{\!J}W_k-i\hbar\pa_{\!J}]\cdot 1,
\eeq
or, equivalently,
\beq
\label{v29}
\delta W_{k}=\delta\Psi_{,A}[\pa_{\!J}W_k-i\hbar\pa_{\!J}]
R^A(\pa_{\!J}W_k-i\hbar\pa_{\!J})\cdot 1 .
\eeq

Taking into account that due to the properties of Legendre transform one has
\beq
\delta\Gamma_k=\delta W_k,
\eeq
the gauge dependence of the 2PI effective action is described by the
following equation
\beq
\label{v30}
\delta\Gamma_k=-
\Big(\frac{\delta\Gamma_{k}}{\delta\Phi^A}+
\frac{\delta\Gamma_{k}}{\delta(\hbar \Delta_k^{AB})}
(\hat{\Phi}^{B}-\Phi^B)\Big)R^A({\hat \Phi})
\delta\Psi[{\hat \Phi}]\cdot 1,
\eeq
or, equivalently, by the equation
\beq
\label{v31}
\delta\Gamma_k=\delta\Psi_{,A}[{\hat \Phi}]
R^A({\hat \Phi})\cdot 1,
\eeq
where ${\hat \Phi}$ is defined in (\ref{v18}) - (\ref{v21}).

Therefore the 2PI effective action (\ref{v14}) repeats the property of
the 2PI effective action introduced in \cite{LSh}
namely it does not depend on gauges when calculating
with use of the equations of motion,
\beq
\label{v32}
\delta\Gamma_k\Big|_{\pa_{{\cal F}}\Gamma_k=0}=0.
\eeq
In turn it allows to state that the 2PI effective action ({\ref{v14}}) defined as a solution
to the following functional integro-differential equation,
\beq
\nonumber
&&\qquad\qquad \exp\Big\{\frac{i}{\hbar}\Big(\Gamma_k[\Phi,\Delta_k]-
\frac{\delta\Gamma_k[\Phi,\Delta_k]}{\delta\Delta_k^{AB}}\Delta_k^{AB}\Big)\Big\}
=\\
\label{v33}
&&=\int D\phi\;
\exp\Big\{\frac{i}{\hbar}\Big(S_{FP}[\Phi+\phi]-
\frac{\delta\Gamma_k[\Phi,\Delta_k]}{\delta\Phi^A}\phi^A-
\frac{1}{2}\frac{\delta\Gamma_k[\Phi,\Delta_k]}{\delta(\hbar\Delta_k^{AB})}
\phi^A\phi^B\Big)\Big\},
\eeq
leads to the gauge independent S-matrix due to the equivalence theorem \cite{KT}.
Expanding the action $S_{FP}[\Phi+\phi]$ in Taylor series
with respect to $\phi^A$ one can find the action
$\Gamma_k=\Gamma_k[\Phi,\Delta_k]$ perturbatively,
\beq
\label{v34}
\Gamma_k=\Gamma_k^{(0)}+\hbar \Gamma_k^{(1)}+O(\hbar^2).
\eeq
For the zero-loop approximation, $\Gamma_k^{(0)}$, from (\ref{v32}) it follows
\beq
\label{v35}
\Gamma_k^{(0)}=S_{FP}[\Phi].
\eeq

The one-loop approximation, $\Gamma_k^{(1)}$,  satisfies the functional
Clairaut-type equation
\beq
\label{v36}
&&\Gamma_k^{(1)}-\frac{\delta\Gamma^{(1)}_k}{\delta\Delta_k^{AB}}
\Delta_k^{AB}=
\frac{i}{2} {\rm STr}\ln\Big(S^{''}_{FP}[\Phi]-
\frac{\delta\Gamma^{(1)}_k}{\delta\Delta_k}\Big),\\
\label{v37}
&&S^{''}_{FP}[\Phi]=\{S^{''}_{FP}[\Phi]_{AB}\},\quad
S^{''}_{FP}[\Phi]_{AB}=\frac{\delta^2S_{FP}[\Phi]}{\delta\Phi^B\delta\Phi^A}.
\eeq
Singular solutions to the ordinary and functional Clairaut-type equations
have been studied in  papers \cite{LMerz,LMerz1}. For the type of equation (\ref{v36})
it was found that the solution can be presented in the form
\beq
\label{v38}
\Gamma_k^{(1)}[\Phi,\Delta_k]=S^{''}_{FP}[\Phi]_{AB}\Delta_k^{AB}-
\frac{i}{2}{\rm STr}\ln \Delta_k,
\eeq
up to some constant quantity.

As a general conclusion, in the perturbation theory
the approach to quantum theory of gauge fields based on concept of
the 2PI effective action is consistent method for describing physical results.
In its turn,  from the beginning the FRG approach is considered as non-perturbative
method to describe quantum properties of gauge fields in terms of
the effective average action found as solutions to the flow equation.
As it was mentioned above the effective average action considering within perturbation
theory is ill-defined because it is gauge dependent functional even on-shell \cite{LSh}.
But for the FRG  it is the more critical  fact  that the effective average action
found as a solution to the flow equation remains ill-defined due to
the gauge dependence at any value of IR parameter $k$ (see  Ref. \cite{Lav1})
that  makes impossible of  physical interpretations of obtained results within this method.
In this connection the study of gauge dependence of the 2PI effective action for the FRG
looks like as very important and actual task.

\section{An alternative flow equation}
\noindent The flow equation in the FRG is the basic relation
describing the dependence of the effective average action on the IR
parameter $k$. Let us derive an alternative flow equation for the
2PI effective action considering firstly the case when external
sources $J$ are free independent variables that do not depend on IR
parameter $k$ and  corresponds to standard approach for composite
operators proposed in \cite{CJT}. Differentiating the functional
$Z_k=Z_k[J,R_k]$ (\ref{v11})
 with respect to $k$ and taking into account that only quantities $R_{k|AB}$
 depend on $k$,  we obtain
\beq
\label{w1}
\pa_k Z_k=\frac{\hbar}{2i}\pa_kR_{k|AB}\;
\frac{\delta^2 Z_k}{\delta J_B\delta J_A},
\eeq
or, equivalently,
\beq
\label{w2}
\pa_k Z_k=\pa_kR_{k|AB}\;\frac{\delta Z_k}{\delta R_{k|AB}}.
\eeq
In terms of the functional $W_k=W_k[J,R_k]$ the equations (\ref{w1}) and (\ref{w2})
rewrite in the form
\beq
\label{w3}
&&\pa_k W_k=\frac{1}{2}\pa_kR_{k|AB}\Big(\frac{\hbar}{i}\frac{\delta^2
W_k}{\delta J_B\delta J_A}+ \frac{\delta W_k}{\delta
J_B}\frac{\delta W_k}{\delta J_A}\Big),\\
\label{w4} &&\pa_k W_k=\pa_kR_{k|AB}\frac{\delta W_k}{\delta
R_{k|AB}}.
\eeq
Due to the Legendre transform (\ref{v15}) variables
$\Phi^A$ are functions of $J_A$ and $R_{k|AB}$ and therefore they
depend on IR parameter $k$ through $R_{k|AB}$. Then we have
\beq
\label{w5} \pa_k\Phi^A=\pa_k\frac{\delta W_k}{\delta
J_A}=\frac{\delta\pa_k W_k}{\delta J_A}= \pa_kR_{k|BC}\frac{\delta^2
W_k}{\delta R_{k|BC}\delta J_A}.
\eeq
In deriving Eq. (\ref{w5}) we
took  into account that the partial derivative $\pa_k$ commutes with
functional derivatives $\delta/\delta J_A$ because variables $J_A$
do not depend on $k$.

Due to properties of the Legendre transform
one has
\beq
\label{w6} \pa_k \Gamma_k=\pa_k W_k,\qquad
\Gamma_k=\Gamma_k[\Phi,\Delta_k].
\eeq
Then from Eqs. (\ref{w4}),
(\ref{w6}), (\ref{v15}), (\ref{v16}) we derive the alternative flow
equation for the 2PI effective action in the FRG,
\beq
\label{w7}
\pa_k \Gamma_k=-
\Big(\pa_k\frac{\delta\Gamma_k}{\delta(\hbar\Delta_k^{AB})}\Big)
\Big(\frac{1}{2}\Phi^B\Phi^A+\hbar\Delta_k^{BA}\Big).
\eeq

Now we are going to consider an another approach to an alternative flow equation
proposed in \cite{AMNS-1}. The main difference is related to requirement for fields $\Phi$
appearing in the Legendre transform to be independent variables on $k$. In turn it leads
to dependence of sources $J$ on $k$. Therefore we have to deal with the following
generating functional of Green functions, ${\bar Z}_{k}[J_k,R_k]$,
and of connected Green functions,
${\bar W}_{k}[J_k,R_k]$,
\beq
\label{w8}
{\bar Z}_{k}[J_k,R_k]=\int
D\phi\;\exp\Big\{\frac{i}{\hbar}\Big[S_{FP}[\phi]+J_{k|A}\phi^A+
\frac{1}{2}R_{k|AB}\phi^B\phi^A \Big]\Big\}=
\exp\Big\{\frac{i}{\hbar}{\bar W}_{k}[J_k,
R_k]\Big\}.
\eeq
Alternative flow equations for functionals ${\bar Z}_{k}[J_k,R_k]$ and
${\bar W}_{k}[J_k,R_k]$ read
\beq
\label{w9}
&&\pa_k {\bar Z}_{k}[J_k,R_k]=\pa_kJ_{k|A}\frac{\delta {\bar Z}_k}{\delta J_{k|A}}+
\pa_kR_{k|AB}\frac{\delta {\bar Z}_k}{\delta R_{k|AB}}, \\
\label{w10}
&&\pa_k {\bar W}_{k}[J_k,R_k]=\pa_kJ_{k|A}\frac{\delta {\bar W}_k}{\delta J_{k|A}}+
\pa_kR_{k|AB}\frac{\delta {\bar W}_k}{\delta R_{k|AB}}.
\eeq
Due to fact that $J_k$ and $R_{k}$ do not depend on gauges
and taking into account
the equations (\ref{v25}) and
(\ref{v28}), the gauge dependence of ${\bar Z}_{k}[J_k,R_k]$ and ${\bar W}_{k}[J_k,R_k]$
is described by the equations
\beq
\label{w11}
\delta {\bar Z}_{k}=\frac{i}{\hbar}\delta\Psi_{,A}[-i\hbar\pa_{\!J_k}]
R^A(-i\hbar\pa_{\!J_k}){\bar Z}_{k},
\eeq
and
\beq
\label{w12}
\delta {\bar W}_{k}=\delta\Psi_{,A}[\pa_{\!J_k}{\bar W}_k-i\hbar\pa_{\!J_k}]
R^A(\pa_{\!J_k}{\bar W}_k-i\hbar\pa_{\!J_k})\cdot 1 .
\eeq
In terms of the 2PI average effective action,
${\bar \Gamma}_k={\bar\Gamma}_k[\Phi, \Delta_k]$,
\beq
\label{w13}
{\bar\Gamma}_k={\bar W}_k-J_{k|A}\Phi^A-R_{k|AB}
\big(\frac{1}{2}\Phi^B\Phi^A+\hbar\Delta_k^{BA}\big),
\eeq
the equation (\ref{w10}) rewrites as
\beq
\label{w14}
\pa_k{\bar\Gamma}_k=-\Big(\pa_k\frac{\delta{\bar\Gamma}_k}{\delta\Phi^A}\Big)\Phi^A+
\Big(\pa_k\frac{\delta{\bar\Gamma}_k}{\delta(\hbar\Delta_{k}^{AB})}\Big)
\big(\frac{1}{2}\Phi^B\Phi^A-\hbar \Delta_{k}^{BA}\big).
\eeq
In deriving the equation (\ref{w14}) the relations $\pa_k\Phi^A=0$
have been taken into account.

In its turn the gauge dependence of ${\bar\Gamma}_k$ according to (\ref{w12})
and $\delta{\bar\Gamma}_k=\delta{\bar W}_k$ is described
by the equation
\beq
\label{w15}
\delta{\bar\Gamma}_k=\delta\Psi_{,A}[{\hat \Phi}]
R^A({\hat \Phi})\cdot 1,
\eeq
where ${\hat \Phi}^A$ are defined by relations similar to (\ref{v18}) -(\ref{v21}).

\section{Gauge dependence of alternative flow equations}
\noindent
Now we are going to investigate the gauge dependence of the flow equations (\ref{w7}) and
(\ref{w14}). Let us consider the case when external sources $J$ do not depend on IR parameter
$k$.
The variation of gauge fixing functional, $\delta\Psi[\phi]$,
does not touch upon a $k$-dependence
of the functional $Z_k=Z_k[J,R_k]$. It allows us to derive the equation describing the gauge
dependence of the flow equation for functional $Z_k$ as
\beq
\label{z1}
\delta\pa_k Z_k=\pa_kR_{k|AB}\;
\frac{\delta(\delta Z_k)}{\delta R_{k|AB}}.
\eeq
Using the equation (\ref{v25}) the equation (\ref{z1}) can be presented in the form
\beq
\label{z2}
\delta\pa_k Z_k=
\frac{i}{\hbar}\pa_kR_{k|AB}\;\delta\Psi_{,C}[-i\hbar\pa_{\!J}]
R^C(-i\hbar\pa_{\!J})\frac{\delta Z_k}{\delta R_{k|AB}}.
\eeq
Variation of the flow equation for functional $W_k=W_k[J,R_k]$ reads
\beq
\label{z3}
\delta\pa_k W_k=
\Big(\pa_kR_{k|AB}\;\delta\Psi_{,C}[\pa_{\!J}W_k-i\hbar\pa_{\!J}]
R^C(\pa_{\!J}W_k-i\hbar\pa_{\!J})\frac{\delta W_k}{\delta R_{k|AB}}
-\pa_k W_k\delta W_k\Big).
\eeq
Taking into account Eq. (\ref{v29}) we obtain the presentation of (\ref{z3})
\beq
\label{z4}
\delta\pa_k W_k=
\frac{i}{\hbar}\pa_kR_{k|AB}\Big[\delta\Psi_{,C}[\pa_{\!J}W_k-i\hbar\pa_{\!J}]
R^C(\pa_{\!J}W_k-i\hbar\pa_{\!J}), \frac{\delta W_k}{\delta R_{k|AB}}\Big]\cdot 1
\eeq
containing the commutator of $\delta\Psi_{,C}[\pa_{\!J}W_k-i\hbar\pa_{\!J}]
R^C(\pa_{\!J}W_k-i\hbar\pa_{\!J})$  and $\delta W_k/\delta R_{k|AB}$.
According to properties of the Legendre transform one has
\beq
\label{z5}
\delta\pa_k \Gamma_k=\delta\pa_k W_k.
\eeq
Therefore the gauge dependence of alternative flow equation for the 2PI effective action
can be described by the following equation
\beq
\label{z6}
\delta\pa_k \Gamma_k=-\frac{i}{\hbar}
\Big(\pa_k\frac{\delta\Gamma_k}{\delta(\hbar\Delta_k^{AB})}\Big)
\Big[\delta\Psi_{,C}[{\hat \Phi}]
R^C({\hat \Phi}), \frac{1}{2}\Phi^B\Phi^A+\hbar\Delta_k^{BA}\Big]\cdot 1 ,
\eeq
where the operators ${\hat \Phi}^A$ are defined in (\ref{v18})-(\ref{v23}).

Now let us study the gauge dependence of alternative flow equation when the external
sources depend on IR parameter $k$, $J_k$, but the fields $\Phi$ appearing in the process
of Legendre transform remain $k$-independent. As to the third possibility
when sources and fields
are $k$-dependent we remain outside our consideration.
Taking into account  arguments used above we arrive
at the equation describing the gauge dependence of
alternative flow equation for functional ${\bar Z}_k$
\beq
\nonumber
\delta\pa_k {\bar Z}_k&=&\frac{i}{\hbar}
\pa_kJ_{k|A}\delta\Psi_{,C}[-i\hbar\pa_{\!J_k}]
R^C(-i\hbar\pa_{\!J_k})\frac{\delta {\bar Z}_k}{\delta J_{k|A}}+\\
\label{z7}
&&
+\frac{i}{\hbar}\pa_kR_{k|AB}\delta\Psi_{,C}[-i\hbar\pa_{\!J_k}]
R^C(-i\hbar\pa_{\!J_k})\frac{\delta {\bar Z}_k}{\delta R_{k|AB}}.
\eeq
In turn the gauge dependence of alternative flow equation for functional ${\bar W}_k$
is ruled by the equation
\beq
\nonumber
\delta\pa_k {\bar W}_k&=&\frac{i}{\hbar}\pa_kJ_{k|A}
\Big[\delta\Psi_{,C}[\pa_{\!J_k}{\bar W}_k-i\hbar\pa_{\!J_k}]
R^C(\pa_{\!J_k}{\bar W}_k-i\hbar\pa_{\!J_k}),
\frac{\delta {\bar W}_k}{\delta J_{k|A}}\Big]\cdot 1+\\
\label{z8}
&&+\frac{i}{\hbar}\pa_kR_{k|AB}\Big[\delta\Psi_{,C}[\pa_{\!J_k}{\bar W}_k-i\hbar\pa_{\!J_k}]
R^C(\pa_{\!J_k}{\bar W}_k-i\hbar\pa_{\!J_k}),
\frac{\delta {\bar W}_k}{\delta R_{k|AB}}\Big]\cdot 1 .
\eeq
Finally the gauge dependence of alternative flow equation for the
2IP average effective action ${\bar \Gamma}_k$
is described by the following equation
\beq
\nonumber
\delta\pa_k {\bar \Gamma}_k&=&-\frac{i}{\hbar}
\Big\{\Big(\pa_k\frac{\delta{\bar \Gamma}_k}{\delta\Phi^A}\Big)-
\Big(\pa_k\frac{\delta{\bar \Gamma}_k}{\delta(\hbar \Delta_k^{AB})}\Big)\Phi^B\Big\}
\big[\delta\Psi[{\hat\Phi}]_CR^C({\hat\Phi}),\Phi^A]\cdot 1-\\
\label{z9}
&&-\frac{i}{\hbar}
\Big(\pa_k\frac{\delta{\bar \Gamma}_k}{\delta(\hbar \Delta_k^{AB})}\Big)
\big[\delta\Psi[{\hat\Phi}]_CR^C({\hat\Phi}),\frac{1}{2}\Phi^B\Phi^A+
\hbar\Delta_k^{BA}]\cdot 1 .
\eeq

We conclude that flow equations for functionals $Z_k$, $W_K$, $\Gamma_k$
and ${\bar Z}_k$, ${\bar W}_K$, ${\bar \Gamma}_k$
depend on gauges
at any finite value of IR parameter $k$. The same statement is valid for solutions to these
equations as well. As to the case when $k\rightarrow 0$ the arguments given in \cite{Lav1}
allow us to confirm the gauge dependence of the 2PI effective action
at the fixed point too.

\section{Summary}
\noindent
In the paper we have analyzed a reformulation of the FRG approach based on using instead
of the effective average action (the 1PI effective action) \cite{Wet1,Wet2}
the 2PI effective action or, in another words,
the effective action with composite operators proposed by Cornwall, Jackiw and
Tomboulis \cite{CJT}. Application of standard FRG approach to gauge systems meets with
serious problem of gauge dependence of the effective average action
even on-shell \cite{LSh}. The first attempt to improve
the situation with gauge dependence in the FRG was made in paper \cite{LSh}
where it was proposed a reformulation of the method \cite{Wet1,Wet2}
with the help of the effective action with composite operators
which were defined as  regulator action densities.
It was shown that in contrast with the effective average action
this effective action within the perturbation theory obeys the important property
of gauge independence when it  is calculated  with using the equations of
motion in full agreement with general statement about gauge dependence
of effective action with composite operators in gauge theories \cite{Lav2}.
In turn,  thanks to the equivalence theorem \cite{KT} it leads
to gauge independence of S-matrix.

But the FRG approach by itself has been introduced as a non-perturbative method to study
quantum properties of field theories. The effective average action should be found
as non-perturbative solution to the flow equation which controls the dependence of
effective average action on the  IR parameter $k$. Quite recently it has been proved
the gauge dependence of the effective average action at any scale of $k$ \cite{Lav1} when
the FRG approach is applied to gauge theories.
It means that in the case of gauge theories the standard FRG approach has no
physical meaning because all obtained
results within this method depend on gauges. For the first sight it seems
that the reformulations of the FRG with the help of the 2PI effective
action given in \cite{LSh,AMNS-1} are more suitable due to good properties
of these approaches in the perturbation theory. Unfortunately,
this expectation does not come true.

We have derived the alternative flow equations in the framework of
standard approach \cite{CJT} when external sources to fields do not depend
on the IR parameter $k$ as well as in the approach \cite{AMNS-1}
when the external sources depend on $k$.
The equation describing the gauge dependence of the
alternative flow equations in both cases has been found.
Analysis of this equation leads to conclusion
that the 2PI effective actions are gauge dependent in any scale of the IR parameter $k$.
Therefore the FRG approach based on the effective average action \cite{Wet1,Wet2}
or on the 2PI effective actions \cite{LSh,AMNS-1}
cannot be considered beyond the perturbation theory
as quantization scheme of gauge fields having physical meaning.

At the moment the last hope to have consistent non-perturbative quantization
procedure of gauge fields  is connected with the BRST exact renormalization group
\cite{Morris1} where
only the absence of an explicit procedure for constructing
a regularized gauge invariant initial action  prevents us
from talking about the completeness of this method \cite{Lav2}.

\section*{Acknowledgments}
\noindent
The work  is supported  by the RFBR grant 18-02-00153 and
by Ministry of Science and High Education of Russian Federation,
project FEWF-2020-0003.


\begin {thebibliography}{99}
\addtolength{\itemsep}{-8pt}

\bibitem{Lav1}
P.M. Lavrov, {\it BRST, Ward identities, gauge dependence and FRG},
arXiv:2002.05997 [hep-th].

\bibitem{Wet1}
C. Wetterich, {\it Average action and the renormalization group
equation}, Nucl. Phys.  {\bf B352} (1991) 529.

\bibitem{Wet2}
C. Wetterich, {\it Exact evolution equation for the effective
potential}, Phys. Lett. {\bf B301} (1993) 90.

\bibitem{brs1}
C. Becchi, A. Rouet, R. Stora, {\it The abelian Higgs Kibble Model,
unitarity of the $S$-operator}, Phys. Lett.  {\bf B52} (1974)
344.

\bibitem{t}
I.V. Tyutin, {\it Gauge invariance in field theory and statistical
physics in operator formalism}, Lebedev Institute preprint  No.  39
(1975), arXiv:0812.0580 [hep-th].

\bibitem{FP}
L.D. Faddeev, V.N. Popov,
{\it Feynman diagrams for the Yang-Mills field},
Phys. Lett. {\bf B25} (1967) 29.

\bibitem{Morris1}
T.R. Morris, {\it Quantum gravity, renormalizability and
diffeomorphism invariance}, SciPost Phys. {\bf 5} (2018) 040,
arXiv:1806.02206[hep-th].

\bibitem{IIS}
Y. Igarashi, K. Itoh, H. Sonoda,
{\it Realization  of  Symmetry  in  the  ERG Approach
to Quantum Field Theory}, Prog.
Theor. Phys. Suppl. {\bf 181} (2009) 1.

\bibitem{Morris2}
Y. Igarashi, K. Itoh, T.R. Morris, {\it BRST in the exact renormalization
group}, Prog. Theor. Exp. Phys. (2019),
arXiv:1904.08231[hep-th].


\bibitem{Lav2}
P.M. Lavrov, {\it RG and BV-formalism},
Phys.Lett. {\bf B803} (2020) 135314.

\bibitem{LSh}
P.M. Lavrov, I.L. Shapiro,
{\it On the functional renormalization group approach for Yang-Mills fields}
JHEP {\bf 1306} (2013) 086.

\bibitem{CJT}
J.M. Cornwall, R. Jackiw, E. Tomboulis,
{\it Effective action for composite operators},
Phys. Rev. {\bf D10} (1974)2428.

\bibitem{Lav-tmph}
P.M. Lavrov, {\it Effective action for composite fields in
gauge theories},
Theor. Math. Phys. {\bf 82} (1990) 282.

\bibitem{AMNS-1}
E. Alexander, P. Millington, J. Nursey, P.M. Safin,
{\it An alternative flow equation for the functional renormalization group},
Phys. Rev. {\bf D100} (2019) 101702.

\bibitem{DeWitt} B.S. DeWitt,
{\it Dynamical theory of groups and fields},
(Gordon and Breach, 1965).

\bibitem{KT}
R.E. Kallosh, I.V. Tyutin, {\it The equivalence theorem and gauge
invariance in renormalizable theories}, Sov. J. Nucl. Phys. {\bf 17}
(1973) 98.

\bibitem{LMerz}
P.M. Lavrov, B.S. Merzlikin,
{\it Loop expansion of average effective action in functional
renormalization group approach},
Phys. Rev. {\bf D92} (2015) 085038.

\bibitem{LMerz1}
P.M. Lavrov, B.S. Merzlikin,
{\it Legendre transformations and Clairaut-type equations},
Phys. Lett.  {\bf B756} (2016) 188.


\end{thebibliography}
\end{document}